\documentclass[aip,jcp,reprint,amsmath,amssymb,floatfix]{revtex4-2}
\usepackage[utf8]{inputenc}
\usepackage{graphicx}
\usepackage{placeins}
\usepackage[capitalize]{cleveref}
\crefname{figure}{Figure}{Figures}
\Crefname{figure}{Figure}{Figures}
\crefname{table}{Tab.}{Tabs.}
\Crefname{table}{Table}{Tables}
\crefname{equation}{Eq.}{Eqs.}
\Crefname{equation}{Equation}{Equations}
\crefname{section}{Sec.}{Secs.}
\Crefname{section}{Section}{Sections}

\begin{document}

\title{Structured High-Angular-Momentum Coulomb Tensors from Real and Complex Solid-Harmonic Integral Engines: A Perspective}

\author{Bo Peng}
\email{peng398@pnnl.gov}
\affiliation{Integrated Discovery Sciences Directorate, Pacific Northwest National Laboratory, Richland, WA 99354 USA}

\begin{abstract}
Electron-repulsion integrals describe the Coulomb interaction between charge distributions built from orbital basis functions.  Most integral algorithms generate these quantities through Cartesian Gaussian functions, whose angular shapes are written as powers of $x$, $y$, and $z$, and then transform the result to spherical functions.  This route is effective, but from $d$ shells onward the Cartesian representation contains more functions than the spherical space required by the calculation.  Direct real or complex solid-harmonic engines work in that target space from the beginning.  They therefore produce a smaller final Coulomb tensor while preserving the ordering, phase, and magnetic-quantum-number labels that describe its angular structure.  Following this structure beyond integral evaluation reveals direct connections to the algorithms that use the tensor.  Simple analytical counts quantify tensor size, angular blocks, radial Slater--Condon parameters, and pair-space work.  These quantities guide low-rank factorization, local Hamiltonian construction, quantum simulation, and transformations to spinor or effective-model bases.  In this way, solid-harmonic integral engines provide a direct bridge between efficient integral generation and structured many-electron computation.
\end{abstract}


\maketitle

\section{Introduction}
\label{sec:introduction}

\begin{figure*}[!tbp]
\centering
\includegraphics[width=0.98\textwidth]{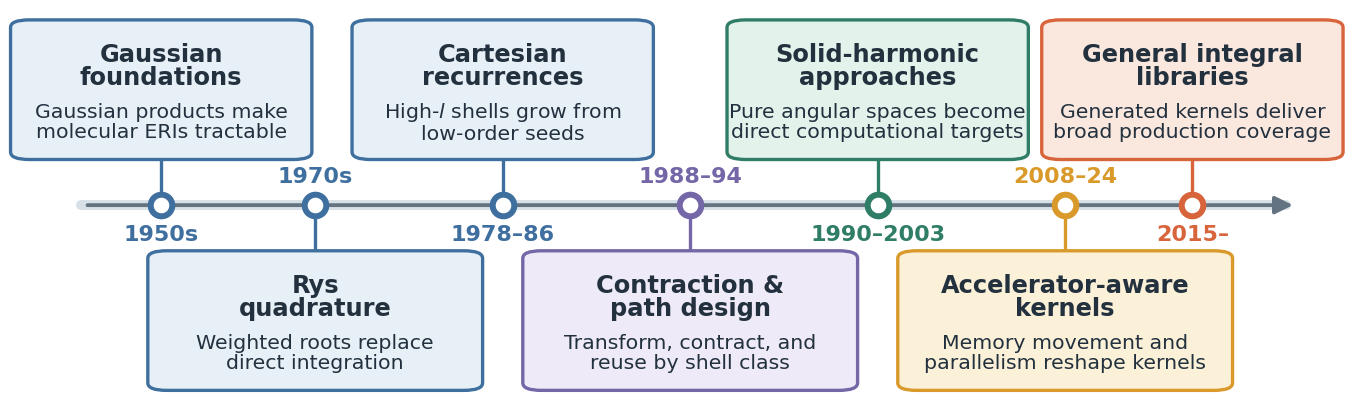}
\caption{Seven decades of ERI design, from tractability to representation awareness.  Successive methods redistributed numerical work, reusable quantities, angular handling, and hardware constraints across integral evaluation.  These method families now coexist, while the direct solid-harmonic route makes angular representation explicit for downstream tensor algorithms.}
\label{fig:integral_timeline}
\end{figure*}

Electronic-structure theory tries to predict how electrons behave in atoms, molecules, and materials.  One of its central physical ingredients is Coulomb repulsion.  Two electrons repel one another more strongly when they are close together, and in atomic units this interaction has the form $1/r_{12}$.  The symbol $r_{12}$ is simply the distance between electron 1 and electron 2.  The physical expression is compact, but evaluating it accurately and repeatedly in a molecular calculation is a major computational task.

To make that task manageable, an electronic orbital is expanded in a finite set of basis functions.  A basis function is a mathematical building block that describes part of the spatial shape of an orbital.  Gaussian basis functions are widely used because products of Gaussians on different atoms can be rewritten as a new Gaussian centered between them.\cite{Boys50_542}  This Gaussian product property converts difficult molecular integrals into quantities that can be built from simpler intermediates.  Each Gaussian basis function also has an angular part.  That angular part determines whether the function has an $s$, $p$, $d$, $f$, or higher-angular-momentum shape around its center.

The electron-repulsion integral, abbreviated ERI, is the central quantity.  An ERI measures the Coulomb interaction between two charge distributions.  Each charge distribution is formed by multiplying two basis functions at the same electron coordinate.  Because there are two basis functions for electron 1 and two for electron 2, an ERI carries four basis-function labels.  In generalized chemists' notation, it is\cite{Boys50_542,Shavitt63_book,Hehre78_161,Davidson78_218,Gill94_141,Sun15_1664}
\begin{equation}
(\mu\nu|\lambda\sigma)
=
\int\!\!\int
\frac{\chi_\mu^*(\mathbf r_1)\chi_\nu(\mathbf r_1)
\chi_\lambda^*(\mathbf r_2)\chi_\sigma(\mathbf r_2)}
{r_{12}}
d\mathbf r_1 d\mathbf r_2.
\label{eq:chemist-eri}
\end{equation}
Here $\chi_\mu$, $\chi_\nu$, $\chi_\lambda$, and $\chi_\sigma$ are basis functions.  The variables $\mathbf r_1$ and $\mathbf r_2$ are the electron positions.  The star denotes complex conjugation.  It has no effect when the basis functions are real, which is the usual situation for nonrelativistic real-spherical calculations.  If the basis contains $N$ functions, the full dense ERI array has a formal size proportional to $N^4$ because each of its four labels can take $N$ values.

Integral algorithms often group the labels in \cref{eq:chemist-eri} as two charge distributions.  Later applications, especially second-quantized Hamiltonians and transformations to complex orbitals, instead group them as the initial and final states of two electrons.  The corresponding two-particle matrix element is therefore defined as
\begin{equation}
V_{\mu\nu;\lambda\sigma}
\equiv
\langle \mu\nu|r_{12}^{-1}|\lambda\sigma\rangle
=
(\mu\lambda|\nu\sigma).
\label{eq:pair-convention}
\end{equation}
The first two labels of $V$ identify the two functions on the bra side, which describes the final two-electron state.  The last two labels identify the ket side, which describes the initial state.  The equality to $(\mu\lambda|\nu\sigma)$ shows exactly how this bra-ket order is related to chemists' notation.  This mapping will later keep the magnetic-quantum-number rule, the second-quantized Hamiltonian, and the spinor transformation consistent.

The word ``quartet'' describes another common grouping.  A basis-function quartet is one choice of four functions, such as $(\mu\nu|\lambda\sigma)$.  A shell quartet is a collection of many such integrals.  A shell contains basis functions on the same atomic center with the same angular momentum.  For example, a real spherical $d$ shell contains five angular functions.  A shell quartet contains every ERI formed by choosing one function from each of four shells.  Integral engines work with shell quartets because many intermediate quantities can then be reused.\cite{Davidson78_218,Saika86_3963,Gill94_141,Sun15_1664}

The history of ERI algorithms is largely the history of making these shell quartets affordable.  Gaussian product formulas first made molecular integrals tractable.\cite{Boys50_542,Shavitt63_book,Saunders75_book,Saunders83_book,Velde84_book,Hegarty83_1135}  Quadrature methods later rewrote an ERI as a weighted sum over numerical roots.\cite{King76_111,Dupuis76_44,King83_154}  Cartesian recurrence methods built high-angular-momentum integrals step by step from simpler starting values.\cite{Davidson78_218,Saika86_3963}  Contraction-aware algorithms then optimized when intermediate quantities should be transformed, combined, and reused.\cite{Pople88_5777,Pople90_5564,Pople91_753,Gill94_141}  Direct solid-harmonic methods asked whether pure spherical angular structure could enter the calculation earlier.\cite{Dunlap90_1127,Dunlap02_032502,Dunlap03_1036,Salvetti93_257,Ishida98_881,Ishida99_4913,Ishida00_7818,Ishida02_378}  More recent libraries and accelerator implementations have emphasized broad integral coverage, vectorization, and memory movement.\cite{Sun15_1664,Ufimtsev08_222,Yasuda08_334,Pritchard16_2537,Wu23_13632,Asadchev24_214110,Li25_1459}  \Cref{fig:integral_timeline} summarizes this cumulative development, in which different methods move different parts of the work.

Across this history, angular representation determines the mathematical language used to describe the shape of a basis function.  Cartesian Gaussian functions use powers of $x$, $y$, and $z$.  For example, the six Cartesian $d$ functions can be associated with $x^2$, $y^2$, $z^2$, $xy$, $xz$, and $yz$.  This language is convenient because the three coordinate directions can be handled separately.  Spherical functions instead describe pure angular momentum.\cite{Fieck79_1063,Dunlap90_1127,Dunlap03_1036,Salvetti93_257}  Real spherical functions are often convenient in nonrelativistic molecular calculations because the orbitals and integrals can remain real.  Complex spherical functions carry an explicit magnetic quantum number $m$.  They are therefore natural for atomic multiplets, spin-orbit coupling, and time-reversal-related spinor partners, which are often called Kramers pairs.\cite{Saue20_204104,Reiher15_book}

Cartesian and spherical $s$ shells both contain one function, and both forms of a $p$ shell contain three.  Their dimensions first differ for $d$ functions.  A Cartesian $d$ shell has six functions, while a pure spherical $d$ shell has five, and the difference grows with angular momentum.  As a result, a conventional Cartesian-to-spherical route may build, store, screen, or transform a larger intermediate tensor before projecting it to the smaller spherical target.\cite{Dunlap90_1127,Dunlap03_1036,Salvetti93_257,Ishida98_881,Ishida99_4913,Ishida00_7818,Ishida02_378}  Mature Cartesian implementations can still be highly efficient, yet the angular labels needed later may appear only after the calculation has passed through the larger representation.

That delay matters because ERIs are rarely the final scientific result.  Density fitting, resolution of the identity, Cholesky decomposition, tensor hypercontraction, compound decomposition, double factorization, and related methods all reorganize or compress the same Coulomb information.\cite{Whitten73_4496,Dunlap79_3396,Vahtras93_514,Weigend02_4285,Beebe77_683,Koch03_9481,Hohenstein12_044103,Parrish12_224106,Peng17_4179,Motta21_83,Berry19_208,Lee21_030305}  Quantum-simulation algorithms use the ERIs to build a fermionic Hamiltonian and then map that Hamiltonian to qubits.\cite{Cao19_10856,McClean16_023023,Babbush18_011044,Babbush18_041015,Takeshita20_011004,VonBurg21_033055}  Relativistic and heavy-element calculations often transform the Coulomb tensor to complex orbitals or spinors.\cite{Foldy50_29,Douglas74_89,Hess86_3742,Dyall97_9618,Kutzelnigg05_241102,Ilias07_064102,Liu09_031104,Saue20_204104,Reiher15_book}  Each application needs not only accurate values, but also a clear statement of the tensor order, normalization, and phases.

High-angular-momentum shells make this interface problem especially visible.  Here ``high angular momentum'' means shells beyond $p$.  A $d$ shell has $l=2$, an $f$ shell has $l=3$, a $g$ shell has $l=4$, and an $h$ shell has $l=5$.  Accurate basis sets and heavy-element calculations often include these functions.\cite{Dunning89_1007,Gomes10_369}  A direct real or complex solid-harmonic engine can produce their Coulomb tensors in the compact spherical space and attach the angular labels at the same time.  The important question is therefore not only whether the engine evaluates integrals quickly.  It is also whether it delivers those integrals in a form that downstream algorithms can use without rebuilding their angular meaning.

\begin{figure*}[!tbp]
\centering
\includegraphics[width=0.98\textwidth]{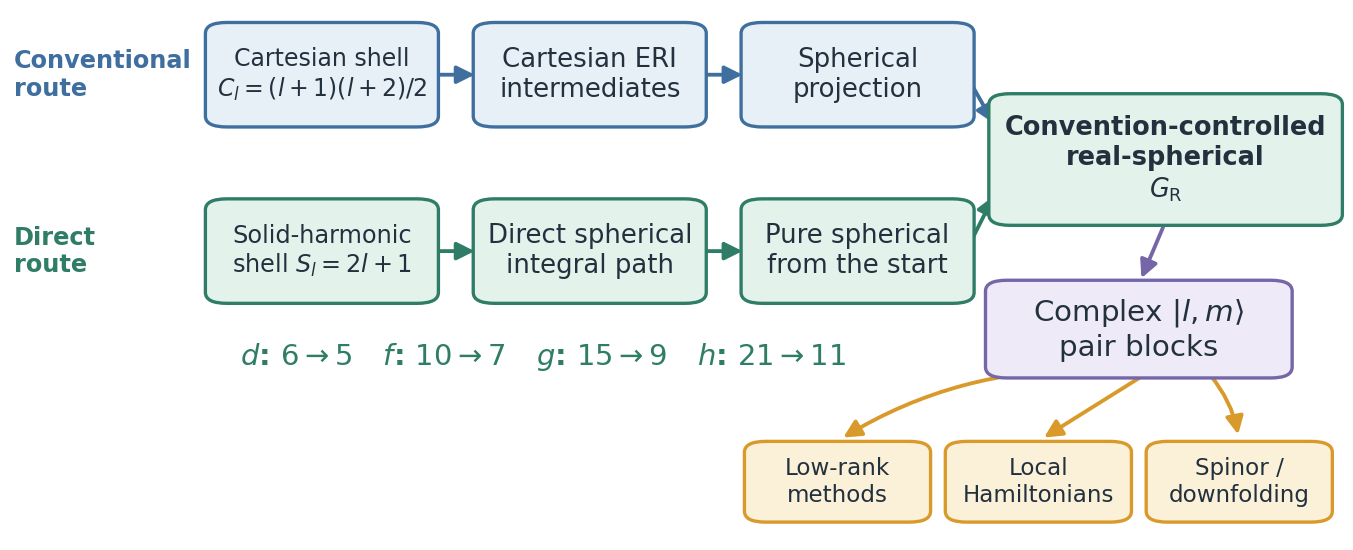}
\caption{Two routes to the same spherical Coulomb tensor.  The difference is when redundant Cartesian angular structure is removed.  The conventional route projects to the real-spherical basis after integral formation, whereas the direct route works in the target shell dimension from the start.  A phase- and ordering-controlled real-to-complex transformation then exposes $|l,m\rangle$ pair blocks for low-rank, local-Hamiltonian, and spinor or downfolding applications.}
\label{fig:representation_workflow}
\end{figure*}

A representation-aware interface connects integral generation to later tensor algorithms.  A solid-harmonic engine can expose convention-controlled spherical shell-pair objects and complex-$m$ blocks, whose analytical structure can then guide low-rank methods, local Hamiltonians, quantum simulation, and spinor transformations.  Exact total-$M$ blocks and the compact Slater--Condon form provide one-center reference limits.  Molecular benchmarks measure how strongly realistic environments depart from those limits.  \Cref{fig:representation_workflow} summarizes the resulting data path.

Understanding this interface begins by separating Cartesian, real-spherical, and complex-spherical representations.  Their dimensions and angular labels lead to exact counts for tensor size, angular blocks, local parameters, and pair-space work.  Those counts then connect integral generation to low-rank preprocessing, local and quantum Hamiltonians, and relativistic or downfolded models.  A common benchmark program can test whether the predicted structure survives in molecules and through downstream transformations.

The broader opportunity is to make angular representation a persistent part of the computational workflow.  Integral libraries and downstream tensor methods could then exchange not only numerical values, but also the angular information needed to interpret and reorganize them.  Establishing that connection across molecular, relativistic, and quantum-simulation benchmarks would turn solid-harmonic integral generation into the starting point for a new class of representation-aware many-electron algorithms.

\section{Why Starting in a Spherical Basis Changes the Tensor}
\label{sec:representation}

The first change is simply the number of functions used to describe one angular-momentum shell.  A Cartesian Gaussian shell with angular momentum $l$ contains\cite{Hehre78_161,Davidson78_218,Dunlap03_1036}
\begin{equation}
C_l=\frac{(l+1)(l+2)}{2}
\end{equation}
functions.  The symbol $C_l$ counts products of $x$, $y$, and $z$ whose powers add to $l$.  For a $d$ shell, examples include $x^2$, $xy$, and $z^2$.  A pure spherical shell instead contains
\begin{equation}
S_l=2l+1
\end{equation}
functions.  Here $S_l$ counts one function for each magnetic label $m=-l,\ldots,l$.  The two counts agree for $s$ and $p$ shells, so the choice of representation does not change their size.  From $d$ shells onward, the Cartesian description contains extra functions.  A $d$ shell has $6$ Cartesian functions but only $5$ spherical functions.  The corresponding counts for $f$, $g$, and $h$ shells are $10$ versus $7$, $15$ versus $9$, and $21$ versus $11$.

An ERI shell quartet has four slots, one for each basis function in the integral.  Therefore, the size difference for one shell is raised to the fourth power for an equal-shell quartet such as $(ll|ll)$.  This fourth-power count isolates the contribution from representation size.  Actual runtime also depends on screening small integrals, combining primitive Gaussians into contracted functions, arranging operations for the processor, and moving data through memory.  A direct spherical calculation nevertheless avoids forming the extra Cartesian entries.

The term ``solid harmonic'' refers to a polynomial with a definite angular momentum.  Multiplying this angular polynomial by a Gaussian factor that controls the radial decay gives a solid-harmonic Gaussian basis function.  A direct solid-harmonic engine uses these functions while it evaluates the integrals.  It therefore avoids first building the larger Cartesian tensor and then projecting that tensor into the spherical basis.\cite{Fieck79_1063,Dunlap90_1127,Dunlap02_032502,Dunlap03_1036,Salvetti93_257}

Once the tensor is spherical, it can be written in either a real or a complex form.  Real spherical functions allow many nonrelativistic molecular calculations to use only real numbers.  Complex spherical functions carry the label $|l,m\rangle$, where $m$ gives the angular momentum projection along a chosen axis.  The two forms contain the same physical information because each real spherical function is a linear combination of complex spherical functions.

Moving safely between the two forms requires three conventions to be stated.  The ordering convention says which function occupies each tensor position.  The normalization convention fixes the scale of each function.  The phase convention fixes its sign or complex phase.  These choices matter because the same $d$ shell can be listed as $(d_{xy},d_{yz},d_{z^2},d_{xz},d_{x^2-y^2})$, as $m=-2,\ldots,2$, or in a library-specific order.  If the choices remain hidden, later transformations to complex spherical functions, spinors, local orbitals, or qubit Hamiltonians can silently permute indices or change signs.

The smaller shell dimension is therefore only the first consequence of starting in a spherical basis.  The explicit angular labels also show which tensor entries are allowed by symmetry, which entries are controlled by the same radial quantities, and which calculations can be separated into independent blocks.  Because these consequences follow from the representation itself, they can be counted before considering the speed of any particular code.

\section{What Can Be Counted Before Timing a Code}
\label{sec:scalings}

Four exact counts separate the effects of spherical representation from the details of a particular implementation.  They describe the number of tensor entries, the entries allowed by angular symmetry, the number of independent radial quantities in an ideal one-center shell, and a rough operation count for calculations organized by magnetic quantum number.  Runtime adds further dependence on the screening threshold, basis contraction, data placement in memory, and hardware.

\subsection{Final Tensor Size}

For an equal-angular-momentum shell quartet $(ll|ll)$, the ratio of spherical tensor entries to Cartesian tensor entries is
\begin{equation}
R_{\mathrm{tensor}}(l)
=
\left(\frac{S_l}{C_l}\right)^4
=
\left[
\frac{2(2l+1)}{(l+1)(l+2)}
\right]^4.
\end{equation}
Here $R_{\mathrm{tensor}}(l)$ compares two entry counts and therefore has no units.  A value of $1$ means that the Cartesian and spherical tensors have the same number of entries.  A value below $1$ means that the spherical tensor is smaller.  The fourth power appears because a two-electron shell quartet has four angular axes.  At large $l$, the ratio approaches $256/l^4$, so the size difference grows rapidly with angular momentum.

The more intuitive quantity is the fractional reduction,
\begin{equation}
\Delta_{\mathrm{tensor}}(l)=1-R_{\mathrm{tensor}}(l).
\end{equation}
For $d$, $f$, $g$, and $h$ equal-shell quartets, this reduction is approximately $51.8\%$, $76.0\%$, $87.0\%$, and $92.5\%$.  These percentages quantify the reduction in stored or processed entries, while timing also reflects which entries an implementation forms and how it evaluates them.

\subsection{Separating the Tensor by Total Magnetic Quantum Number}

The entry count describes the size of the tensor but not its internal organization.  The complex spherical basis reveals a second structure because every function has a magnetic label $m$.  Consider an ideal one-center shell, meaning that all functions share one center and the interaction is unchanged when the system is rotated.  For the scalar Coulomb tensor defined in \cref{eq:pair-convention}, the sum of the two magnetic labels is then conserved between the bra and ket pairs:\cite{Slater29_1293,Condon30_1121,Racah42_438,Fieck79_1063,Dunlap02_032502}
\begin{equation}
m_1+m_2=m_3+m_4.
\label{eq:total-m-selection}
\end{equation}
The labels $m_1$ and $m_2$ belong to the bra pair of $V_{m_1m_2;m_3m_4}$, while $m_3$ and $m_4$ belong to the ket pair.  Their sums define the total magnetic quantum number $M$ of each pair.  Equation~\eqref{eq:total-m-selection} therefore says that the Coulomb interaction connects only pairs with the same $M$.

Let $n=2l+1$ be the number of complex spherical functions in the shell.  A dense four-index local tensor has $n^4$ entries.  The number of entries allowed by the total-$M$ rule is
\begin{align}
N_M(l)
&=
\sum_{M=-2l}^{2l}
\left(2l+1-|M|\right)^2 \notag \\
&=
(2l+1)^2
+2\sum_{M=1}^{2l}(2l+1-M)^2 \notag\\
&=
(2l+1)^2
+2\sum_{j=1}^{2l}j^2 \notag\\
&=
(2l+1)^2
+\frac{(2l)(2l+1)(4l+1)}{3} \notag\\
&=
\frac{(2l+1)(8l^2+8l+3)}{3}.
\label{eq:total-m-count}
\end{align}
The counting follows directly from this rule.  For a fixed value of $M$, there are $n-|M|=2l+1-|M|$ ordered pairs $(m_1,m_2)$.  The bra and ket can each choose any pair with that same value, which gives $(n-|M|)^2$ entries in the $M$ block.  The $M=0$ block contributes $n^2$ entries, while the $+M$ and $-M$ blocks have equal sizes.  These observations give the second line of \cref{eq:total-m-count}.  The remaining lines use the standard formula for the sum of squares.  For example, a $d$ shell has block sizes $1,2,3,4,5,4,3,2,1$.  Squaring and adding these sizes gives $85$, which agrees with \cref{eq:total-m-count} for $l=2$.  At large $l$, the allowed fraction $N_M(l)/n^4$ decreases approximately as $1/(3l)$.

In a molecule, nearby atoms change the local environment.  Their electric fields, orbital mixing, and basis functions on other centers can therefore couple different values of $M$.  The one-center pattern provides a reference against which this mixing can be measured.  Section~\ref{sec:low-rank} introduces such a measurement after first explaining a second kind of one-center structure.

\subsection{Replacing Many Entries with a Few Radial Parameters}

The total-$M$ rule tells us which entries can be nonzero.  Rotational symmetry provides an additional simplification by relating the values of those allowed entries.  The Slater--Condon decomposition expresses these relations through a small set of radial parameters.\cite{Slater29_1293,Condon30_1121,Racah42_438,Racah43_367,Sugano70_book,Griffith61_book}  Each parameter measures a different radial part of the Coulomb interaction, while fixed angular formulas determine how it contributes to the tensor.  Many tensor entries therefore share the same underlying physical quantities.

A rotationally invariant Coulomb interaction within a fixed $l$ shell can be expressed through even-rank radial parameters,
\begin{equation}
k=0,2,\ldots,2l.
\end{equation}
The integer $k$ labels one angular pattern, or channel, of the Coulomb interaction.  Only even $k$ values appear for a scalar Coulomb interaction within a fixed shell.  The number of radial parameters is therefore
\begin{equation}
N_{\mathrm{SC}}(l)=l+1.
\end{equation}
The notation $N_{\mathrm{SC}}$ means the number of Slater--Condon radial parameters.  A $d$ shell uses $F^0$, $F^2$, and $F^4$.  An $f$ shell also uses $F^6$, while $g$ and $h$ shells require five and six parameters.  The parameter $F^0$ sets the repulsion averaged over all directions.  The higher parameters describe how that repulsion changes with direction and how it splits atomic states with different arrangements of the electrons.  These splittings are often called multiplet splittings.  The parameter count is therefore much smaller than the $n^4=(2l+1)^4$ entries of a dense local tensor.

The usual Slater--Condon form can be summarized as
\begin{equation}
V_{m_1m_2;m_3m_4}
=
\sum_{k=0,2,\ldots,2l}
F^k B_{m_1m_2;m_3m_4}^{(k,l)}.
\end{equation}
Here $V_{m_1m_2;m_3m_4}$ is one local Coulomb matrix element in the complex spherical basis.  The parameter $F^k$ gives the radial strength of channel $k$.  The tensor $B^{(k,l)}$ contains fixed numbers that distribute this strength among the angular states.  These numbers include Gaunt coefficients, which are integrals of products of spherical harmonics.\cite{Fieck79_1063,Dunlap02_032502}  The equation therefore separates the radial part that depends on the shell from the angular part fixed by symmetry.

This compact form reconstructs every symmetry-allowed tensor entry from $l+1$ parameters.  In other words, the parameters generate the full tensor rather than selecting $l+1$ of its entries.  The relation is exact when the functions belong to one rotationally symmetric shell and share the same radial shape.  In a molecule or embedded region, the same formula becomes a model that can be fitted to the calculated tensor.  The difference between the fitted model and the tensor then measures how much of the ideal angular structure survives.

\subsection{Organizing Calculations in Pair Space}

\begin{figure*}[!tbp]
\centering
\includegraphics[width=0.98\textwidth]{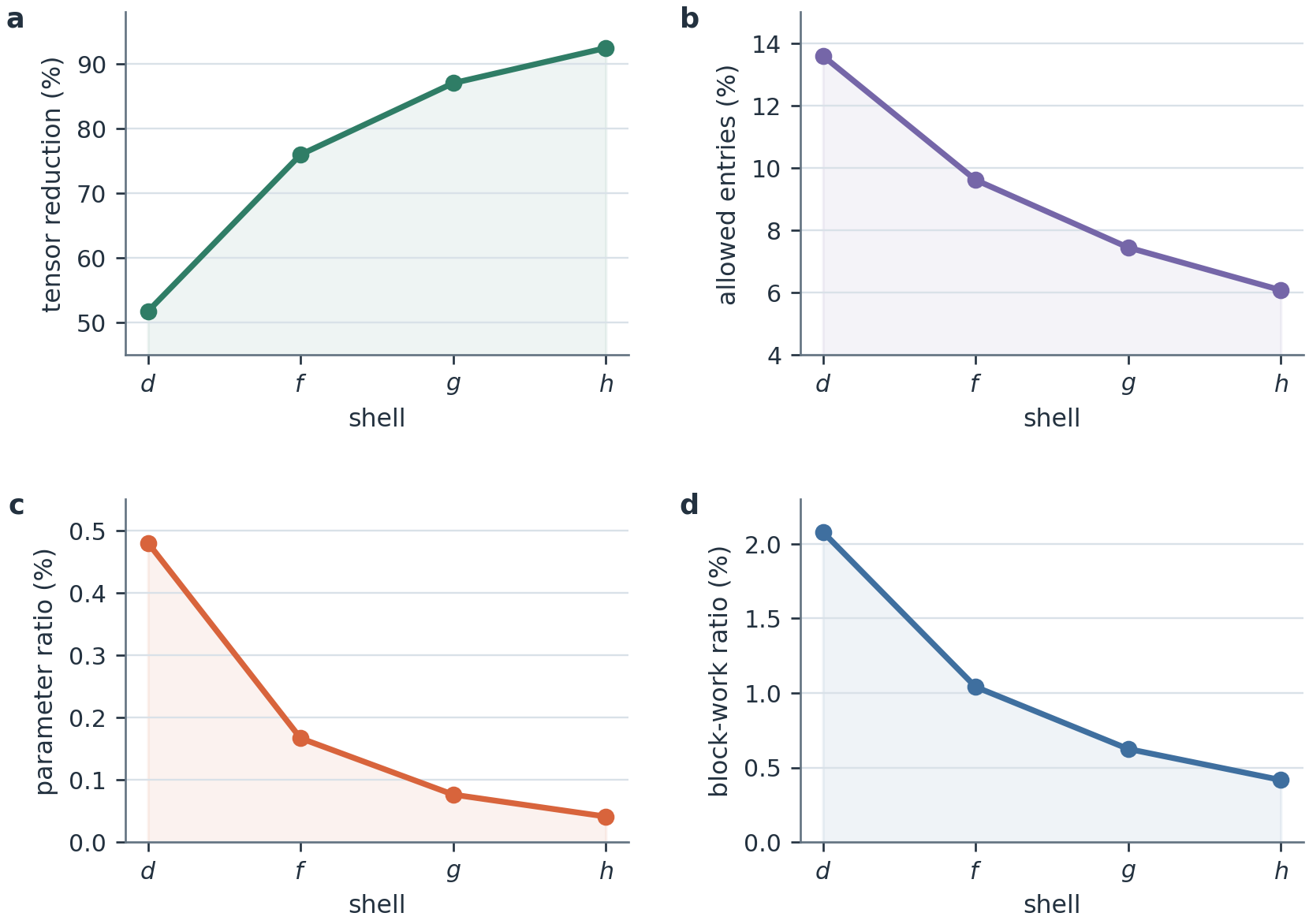}
\caption{Four ratios determined by angular representation for equal-$l$ shells from $d$ through $h$.  Panel (a) shows the percentage reduction in final spherical versus Cartesian tensor entries.  Panel (b) shows the percentage of complex-spherical entries that satisfy the total-$M$ rule.  Panel (c) compares the number of independent Slater--Condon radial parameters with the number of entries in the full spherical tensor.  It counts independent inputs rather than retained tensor entries.  Panel (d) compares a rough operation count for blockwise pair-space updates with the corresponding dense count.  Together, the panels isolate representation effects from implementation-dependent timing.}
\label{fig:analytical_scaling}
\end{figure*}

The previous two counts describe the tensor itself.  The fourth count describes work performed on that tensor.  Many algorithms combine two orbital indices into one pair index and then treat the four-index Coulomb tensor as an ordinary matrix.  The resulting matrix is said to be in pair space.  Density fitting, Cholesky decomposition, and related methods all use this organization.\cite{Whitten73_4496,Dunlap79_3396,Vahtras93_514,Weigend02_4285,Beebe77_683,Koch03_9481,Hohenstein12_044103}  If the shell has $n=2l+1$ functions, then the number of ordered pairs is
\begin{equation}
N_{\mathrm{pair}}=n^2.
\end{equation}
A calculation that combines every pair with every other pair often has a leading operation count proportional to the cube of the pair-space dimension.  A simple cubic reference count is
\begin{equation}
N_{\mathrm{pair}}^3=n^6.
\end{equation}
If the pair states are grouped by total magnetic quantum number $M$, then the block with label $M$ has size
\begin{equation}
n_M=2l+1-|M|.
\end{equation}
The corresponding blockwise work proxy is
\begin{equation}
K_M(l)
=
\sum_{M=-2l}^{2l}n_M^3
=
(2l+1)^2(2l^2+2l+1).
\end{equation}
The ratio $K_M(l)/n^6$ decreases approximately as $1/(8l^2)$ at large $l$.  It compares idealized operation counts for blockwise and dense calculations.  Separating pair states by $M$ may therefore reduce the work of matrix updates, low-rank factorizations, and later reductions that operate block by block.

\Cref{fig:analytical_scaling} shows how these four consequences grow from $d$ through $h$ shells.  The first ratio measures how much smaller the spherical tensor is.  The second and third describe two levels of physical structure within that tensor: the entries allowed by angular momentum and the radial parameters that determine their values.  The fourth shows how an algorithm could use the block structure to organize its work.  This sequence leads from tensor representation to the first application, low-rank factorization.

\section{Application I: Organizing Low-Rank Coulomb Representations}
\label{sec:low-rank}

\begin{figure*}[!tbp]
\centering
\includegraphics[width=0.98\textwidth]{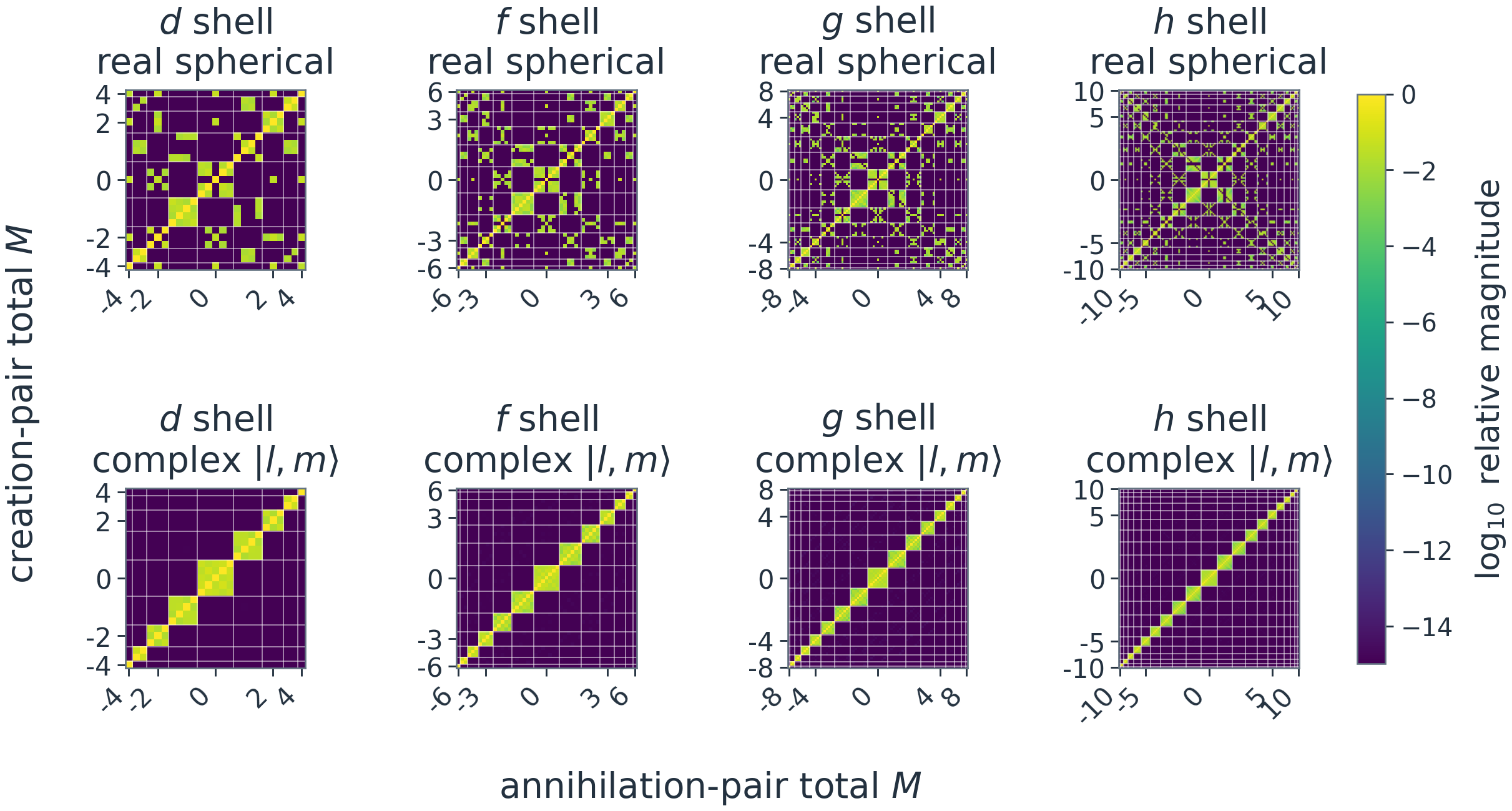}
\caption{Coulomb matrices in pair space for one-center $d$, $f$, $g$, and $h$ shells before and after the reversible real-to-complex spherical transformation.  Pair states are sorted by total magnetic quantum number $M$.  In the real-spherical basis, nonzero entries are spread across the matrices.  In the complex $|l,m\rangle$ basis, they collect into the blocks required by $m_1+m_2=m_3+m_4$.  Color shows the base-10 logarithm of each magnitude relative to the largest entry in that matrix.  In a molecule, coupling between different blocks would appear as measurable intensity outside the bright block pattern.}
\label{fig:total_m_blocks}
\end{figure*}

Low-rank methods reduce storage and computation by writing a large Coulomb tensor as products of smaller arrays.  Density fitting, also called resolution of the identity (RI), introduces an auxiliary basis that approximates products of orbital basis functions.  Cholesky decomposition instead builds the smaller factors directly from the Coulomb matrix in pair space.  Tensor hypercontraction, compound decomposition, double factorization, and factorizations for quantum simulation break the interaction into other sequences of smaller factors.  Although their formulas differ, all of these methods avoid carrying every element of the dense four-index tensor explicitly.\cite{Whitten73_4496,Dunlap79_3396,Vahtras93_514,Weigend02_4285,Beebe77_683,Koch03_9481,Hohenstein12_044103,Parrish12_224106,Peng17_4179,Motta21_83,Berry19_208,Lee21_030305}

All of these methods combine two orbital labels into a pair label.  Consequently, the angular representation becomes part of their basic data structure.  A real spherical pair is compact, but its magnetic labels are not directly visible.  A complex spherical pair carries $m_1$ and $m_2$, so its total label is $M=m_1+m_2$.  In the ideal one-center limit, pairs with different values of $M$ do not interact.  The pair matrix can therefore be separated into smaller independent blocks:
\begin{equation}
V=\bigoplus_M V^{(M)}.
\end{equation}
Here $V$ is the full local Coulomb matrix in pair space, and $V^{(M)}$ contains only the pairs with total magnetic quantum number $M$.  The direct-sum symbol means that the blocks can be stored and processed independently.

The real-to-complex transformation is unitary, meaning that it is a reversible change of basis that preserves lengths and angles.  It therefore preserves the matrix rank, which is the minimum number of independent factors needed to represent the matrix exactly, while making the block structure explicit.  Under exact one-center symmetry, a Cholesky approximation can then be written one block at a time:
\begin{equation}
V_{pq}^{(M)}
\approx
\sum_P L_{pP}^{(M)}L_{qP}^{(M)*}.
\end{equation}
The indices $p$ and $q$ label pair states inside one total-$M$ block.  The index $P$ labels one Cholesky vector, which is one column of the low-rank factor $L^{(M)}$.  Because each factor is confined to a block, a blockwise implementation can avoid storing zeros between different values of $M$.  In a molecule, the blocks mix and the useful numerical sparsity then depends on the chosen orbitals, the error threshold, the order in which factor vectors are selected, and the factorization method.  Direct measurement therefore determines how much block structure remains.

As \cref{fig:analytical_scaling} shows, the fraction of entries inside the ideal total-$M$ blocks decreases with $l$, and the rough blockwise operation count decreases even faster.  High-$l$ shells are therefore useful test cases for RI, Cholesky, and quantum factorizations that can work with blocks.  To test whether the one-center pattern remains useful in a molecule, first construct a reference tensor that keeps only entries with the same $M$ on the bra and ket:
\begin{equation}
\left(V_{\mathrm{block}}\right)_{m_1m_2;m_3m_4}
=
\delta_{m_1+m_2,m_3+m_4}
V_{m_1m_2;m_3m_4}.
\label{eq:block-projector}
\end{equation}
The Kronecker delta in \cref{eq:block-projector} equals one when the two sums of magnetic labels are equal and zero otherwise.  It therefore keeps entries inside the ideal blocks and removes all other entries.  The fraction of the tensor outside those blocks is
\begin{equation}
\eta_{\mathrm{off}}
=
\frac{\|V-V_{\mathrm{block}}\|_F}{\|V\|_F}.
\label{eq:eta-off}
\end{equation}
Here $\|\cdot\|_F$ is the Frobenius norm, which is the square root of the sum of the squared magnitudes of all tensor entries.  A small $\eta_{\mathrm{off}}$ means that most of the tensor remains inside the ideal blocks.  A large value means that the molecular environment or the chosen orbitals strongly mix different values of $M$.

The value of $\eta_{\mathrm{off}}$ depends on which local orbitals are included and how they are defined.  A reproducible benchmark must therefore state how the local orbital subspace was selected and how its orbitals were made orthonormal, meaning mutually perpendicular and individually normalized.  It must also report the local center and axes, the real and complex spherical order, the phase and normalization conventions, and the four tensor axes under \cref{eq:pair-convention}.  If the axes are rotated to minimize $\eta_{\mathrm{off}}$, the benchmark should give both the value in a fixed chemically defined frame and the minimized value.  Reporting these choices makes comparisons across molecules and codes meaningful.

This measurement also suggests a different computational route.  A conventional calculation may first form Cartesian intermediate arrays, project them to spherical functions, assemble a four-index tensor, transform it to complex or local orbitals, and only then begin the factorization.  A representation-aware route could instead form spherical shell-pair objects during integral evaluation and pass those objects directly to RI or Cholesky routines.  When the factorization needs only selected blocks, newly chosen factor columns, or the remaining approximation error, this route could avoid explicitly forming larger intermediate tensors.

\Cref{fig:total_m_blocks} makes the reorganization visible.  The real and complex matrices contain the same information, but the complex basis gathers related pair states into separate blocks.  A molecular calculation can use $\eta_{\mathrm{off}}$ to measure how much intensity moves outside this ideal pattern.

Pair-space blocks organize the numerical calculation.  If the local interaction also remains close to rotationally symmetric, the Slater--Condon relations organize the physical content of the same tensor using only a few radial parameters.  That additional connection leads directly to local Hamiltonians and quantum simulation.

\section{Application II: Building Local Interactions for Quantum Simulation}
\label{sec:local}

\begin{figure*}[!tbp]
\centering
\includegraphics[width=0.98\textwidth]{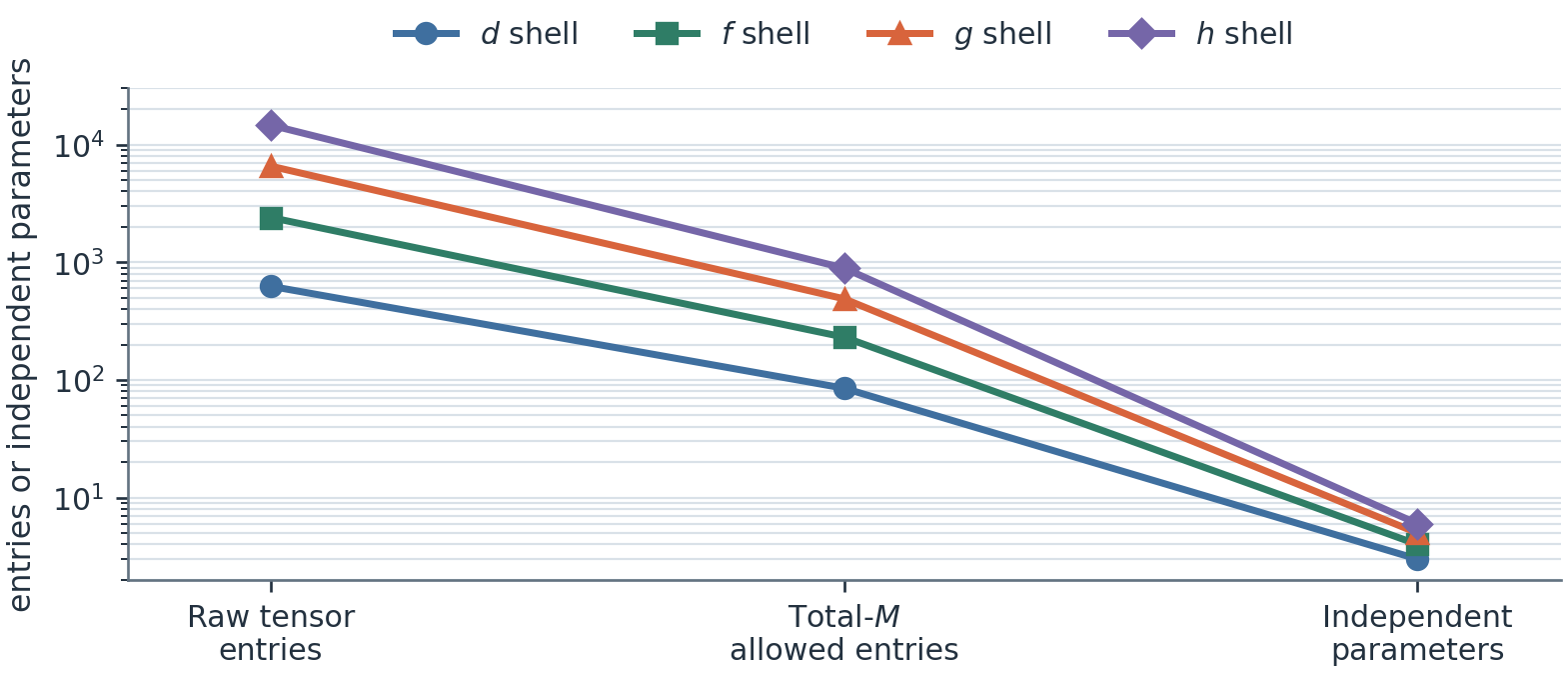}
\caption{Three levels of one-center Coulomb information for $d$, $f$, $g$, and $h$ shells.  Each line begins with every entry in the four-index spherical tensor.  The middle point counts the entries that satisfy the total-$M$ rule.  The final point counts the independent Slater--Condon radial parameters that determine all of those allowed entries in an ideal rotationally symmetric shell.  Thus the three points represent the full tensor, its symmetry-allowed entries, and the smaller set of physical inputs from which those entries can be reconstructed.}
\label{fig:local_compression}
\end{figure*}

Quantum simulation does not begin with qubits.  It begins with a fermionic Hamiltonian, an operator that describes electrons in a chosen set of orbitals and respects their particle statistics.  Creation and annihilation operators in that Hamiltonian add or remove an electron from an orbital.  A Jordan--Wigner, Bravyi--Kitaev, or related transformation then translates these electron operators into operations on qubits.\cite{Jordan28_631,Bravyi02_210,Seeley12_224109,Cao19_10856,McClean16_023023}  Because the translation occurs after the electronic Hamiltonian is built, the structure of the final qubit Hamiltonian depends on the Coulomb tensor supplied at the beginning.

If a local high-$l$ interaction is supplied only as a dense tensor, its many entries can look like unrelated coefficients.  In an atomic-like shell, however, different arrangements of the electrons form related energy levels called multiplets.  The Slater--Condon construction explains these relationships by deriving every tensor entry from a few radial parameters and fixed angular formulas.  The parameter $F^0$ sets the average repulsion, while the higher $F^k$ values describe how that repulsion depends on the angular arrangement of the electrons.\cite{Slater29_1293,Condon30_1121,Racah42_438,Racah43_367,Sugano70_book}

This shared structure becomes useful when an active space contains local $d$ or $f$ shells together with higher-$l$ polarization functions.\cite{Roos80_157,Sayfutyarova17_4063}  An active space is the selected group of orbitals whose electron configurations are treated explicitly.  Polarization functions give those orbitals additional freedom to change shape.  Within this space, a dense tensor can hide which coefficients come from the same angular interaction.  A Slater--Condon-like representation keeps those relationships visible.  It can therefore connect the tensor to model quantities such as Hubbard $U$, which measures an average local repulsion, and Hund coupling, which describes important energy differences between spin arrangements.  Kanamori and Racah parameters provide related ways to describe the same local interaction when their assumptions are appropriate.\cite{Kanamori63_275,Anisimov91_943,Liechtenstein95_R5467,Georges13_137}

Each radial parameter contributes to many electron-interaction terms, and each of those terms can become several Pauli strings, which are products of simple operators acting on individual qubits.  The final qubit Hamiltonian can therefore remain large even though its coefficients originate from a small set of related Coulomb inputs.  Keeping these relationships visible can help organize coefficients, group terms, compare active spaces, and trace errors through the mapping.\cite{Babbush18_011044,Babbush18_041015,Motta21_83,Berry19_208,Takeshita20_011004,VonBurg21_033055}

To preserve these relationships, the tensor convention must agree with the order of the electron operators.  With the convention in \cref{eq:pair-convention}, one fixed-$l$ complex-spherical matrix element is
\begin{align}
V_{m_1m_2;m_3m_4}
&=
\int\!\!\int
\phi_{lm_1}^*(\mathbf r_1)
\phi_{lm_2}^*(\mathbf r_2)
\frac{1}{r_{12}}
\notag\\
&\qquad\times
\phi_{lm_3}(\mathbf r_1)
\phi_{lm_4}(\mathbf r_2)
d\mathbf r_1d\mathbf r_2.
\label{eq:complex-pair-integral}
\end{align}
The same tensor enters the following operator form for a local interaction that preserves spin:
\begin{align}
\hat H_U
=
\frac{1}{2}
\sum_{\substack{\sigma,\tau\\m_1,m_2,m_3,m_4}}
V_{m_1m_2;m_3m_4}
\hat a_{m_1\sigma}^{\dagger}
\hat a_{m_2\tau}^{\dagger}
\hat a_{m_4\tau}
\hat a_{m_3\sigma}.
\end{align}
Here $\sigma$ and $\tau$ label the electron spins.  The operator $\hat a^\dagger$ creates an electron in the orbital named by its subscripts, while $\hat a$ removes one.  The factor of $1/2$ avoids counting the same electron pair twice.  Because \cref{eq:complex-pair-integral} fixes both complex conjugation and index order, the total-$M$ rule and the operator order in $\hat H_U$ refer to the same tensor entries.

A solid-harmonic integral engine can therefore prepare more than a list of Coulomb values.  It can provide local blocks with the angular labels and index order needed to connect the original integrals to multiplet models and, later, to qubit operators.

\Cref{fig:local_compression} shows the two reductions in sequence.  Choosing the complex spherical basis first reveals which entries satisfy the total-$M$ rule.  Rotational symmetry then relates the values of those allowed entries through a much smaller set of radial parameters.

This traceable angular structure becomes even more important when orbital and spin labels are combined.  Relativistic calculations use spinors that mix those two kinds of information, which makes the Coulomb tensor's change of basis central to preserving its physical meaning.

\section{Application III: Transforming Coulomb Tensors for Relativistic and Effective Models}
\label{sec:spinor}

\begin{figure*}[!tbp]
\centering
\includegraphics[width=0.98\textwidth]{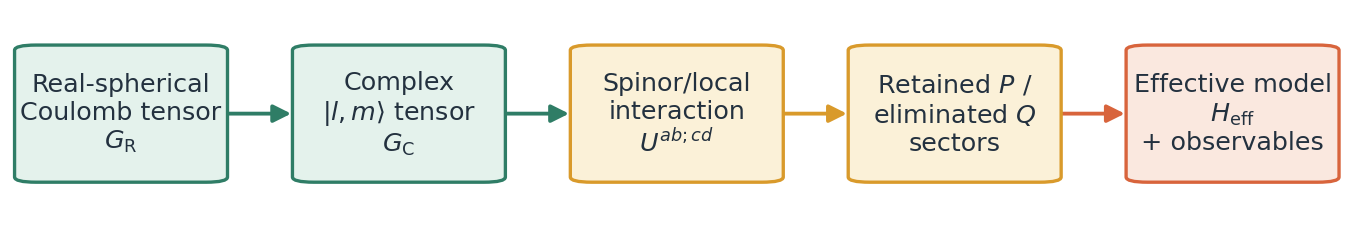}
\caption{A Coulomb tensor passes through several representations on its way to an effective model.  The first transformation converts the real-spherical tensor $G_{\mathrm R}$ to the complex $|l,m\rangle$ tensor $G_{\mathrm C}$.  The next transformation produces interactions in spinor or local orbitals.  The low-energy reduction then separates the states that are kept ($P$) from those removed from the explicit model ($Q$), producing an effective Hamiltonian and transformed observables.  Gaunt, Breit, and other spin-dependent two-electron corrections require additional operators and are not supplied by the scalar Coulomb engine.}
\label{fig:spinor_flow}
\end{figure*}

Relativistic electronic-structure methods introduce another change of representation.  Their working orbitals can be complex spinors, which are functions with coupled spatial and spin components.  Four-component Dirac methods and exact-two-component methods obtain these spinors from a relativistic one-electron equation.  This description is especially important for heavy elements.  There, scalar relativistic effects change orbital energies and shapes, while spin-orbit coupling links an electron's spatial motion to its spin.\cite{Foldy50_29,Douglas74_89,Hess86_3742,Dyall97_9618,Kutzelnigg05_241102,Ilias07_064102,Liu09_031104,Peng13_184105,Saue20_204104,Reiher15_book}

The ordinary scalar Coulomb interaction remains one part of this relativistic problem.  A solid-harmonic Coulomb engine can supply that part with a well-defined ordering and phase convention.  Additional operators are needed for a complete relativistic two-electron treatment.  These include Gaunt and Breit terms, which describe magnetic and retardation corrections, as well as picture-change corrections that arise when operators are transformed between relativistic representations.\cite{Peng13_184105,Saue20_204104,Reiher15_book}  Keeping these contributions separate makes the role of the scalar Coulomb tensor clear.

The scalar tensor can then pass through the following sequence:
\begin{align}
G_{\mathrm R}
&\rightarrow
G_{\mathrm C}(|l,m\rangle)
\rightarrow
U_{\mathrm{spinor/local}}
\notag\\
&\rightarrow
H_{\mathrm{eff}}.
\end{align}
Here $G_{\mathrm R}$ is the Coulomb tensor in a real spherical basis, and $G_{\mathrm C}$ is the same tensor in the complex spherical $|l,m\rangle$ basis.  Both use the bra-ket axis order in \cref{eq:pair-convention}.  Transforming the orbital indices produces $U_{\mathrm{spinor/local}}$, the interaction in a spinor or localized-orbital basis.  A later reduction produces $H_{\mathrm{eff}}$, an effective Hamiltonian for the low-energy states that the calculation keeps.

The orbital transformation can be written schematically as
\begin{equation}
U_i^{ab;cd}
=
\sum_{\mu\nu\lambda\sigma}
C_{\mu a}^{i*}C_{\nu b}^{i*}
V_{\mu\nu;\lambda\sigma}
C_{\lambda c}^{i}C_{\sigma d}^{i}.
\end{equation}
The site label $i$ identifies an atom, fragment, or other region chosen for a local model.  The labels $a$, $b$, $c$, and $d$ identify the local orbitals kept in that model, while the Greek labels identify the original atomic-orbital functions.  The coefficients $C^i$ express each local orbital as a combination of the original functions.  Because the first two axes of $V$ belong to the bra side, the corresponding coefficients are complex conjugated.  The equation therefore carries the tensor convention in \cref{eq:pair-convention} into the local basis.

The next step, often called downfolding, replaces a large Hamiltonian with a smaller one for selected low-energy states.  High-energy states are removed from the explicit calculation, but their influence is retained through modified interactions and observables.  Schrieffer--Wolff transformations, Bloch and Okubo effective Hamiltonians, similarity-renormalization methods, and flow-equation approaches are different ways to carry out this reduction.\cite{Okubo54_603,Bloch58_329,Schrieffer66_491,Glazek93_5863,Glazek94_4214,Wegner94_77,White02_7472,Kehrein06_book}  Because the reduction must update both the interactions and the measured quantities, forming a dense spinor Coulomb tensor beforehand can require substantial storage and transformation work.

A representation-aware interface offers a different order of operations.  A solid-harmonic engine can provide real or complex spherical pair blocks before the calculation forms the full spinor tensor.  Later code can transform, screen, factor, or select those blocks while their angular labels remain visible.  The physical reduction is unchanged, but the calculation gains control over which intermediate tensors are formed and which labels remain available at each step.

This traceability is central for multiorbital Hubbard models with spin-orbit coupling.  In these models, the retained orbitals mix spin and high-$l$ angular character, so a single unlabeled interaction value cannot describe the local physics.\cite{Jackeli09_017205}  The full tensor $U_i^{ab;cd}$ is needed, and each of its axes should remain connected to the parent angular momentum, spinor composition, and multiplet structure.  A convention-controlled solid-harmonic tensor provides that starting point.  \Cref{fig:spinor_flow} separates the changes of basis from the later removal of high-energy states.

Low-rank factorization, local-model construction, and relativistic or low-energy transformations therefore share one requirement: angular information must remain useful and traceable after each change of representation.  A common set of measurable tests can determine whether that requirement is met.

\section{From Perspective to a Testable Benchmark Program}
\label{sec:benchmark}

\begin{figure*}[!tbp]
\centering
\includegraphics[width=0.98\textwidth]{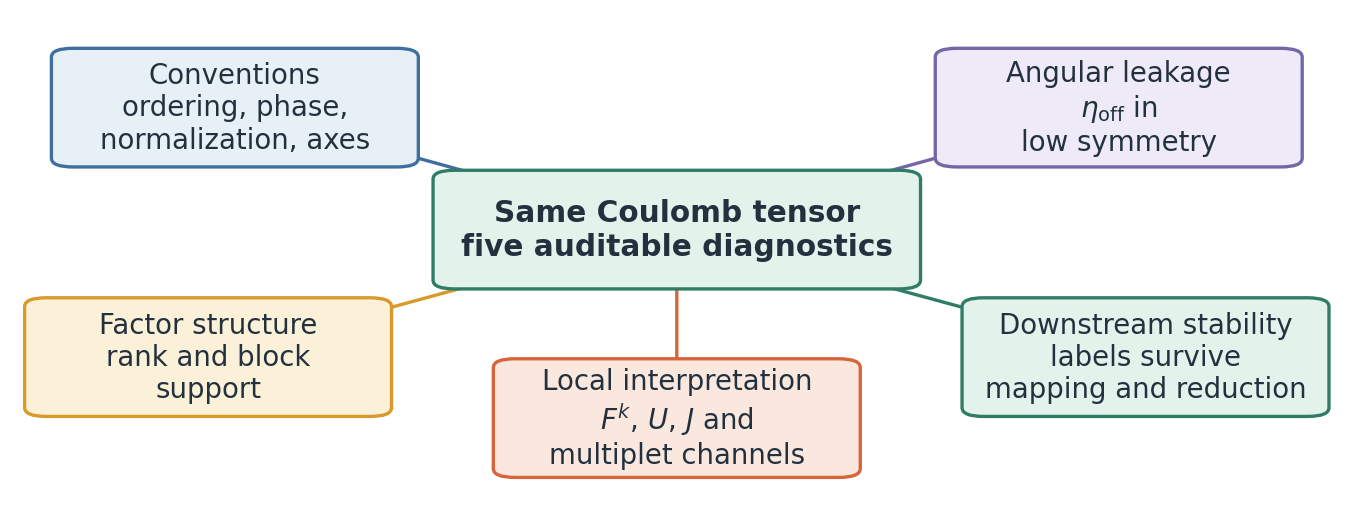}
\caption{Five checks follow the same Coulomb tensor from integral generation to later applications.  First, a complete record of ordering, phase, normalization, and axis conventions makes every transformation reproducible.  Second, the weight outside ideal total-$M$ blocks measures how strongly a molecular environment mixes angular sectors.  Third, the locations of nonzero factor entries show whether the block pattern remains useful in a low-rank representation.  Fourth, fitted local parameters test whether a dense tensor retains a simple physical interpretation.  Fifth, the final calculation checks whether the original angular labels remain traceable after mapping and reduction.}
\label{fig:benchmark_program}
\end{figure*}

The three applications raise two separate questions.  How quickly can an integral engine produce the tensor, and how much of the tensor's angular structure remains useful afterward?  Timings alone answer only the first question.  They also depend on engineering choices such as screening, which skips negligible integrals, batching, which groups similar work, and vectorization, which processes many values together.  A useful benchmark should therefore combine performance measurements with diagnostics, meaning measurable checks of the tensor and its later transformations.  The five checks in \cref{fig:benchmark_program} follow the same Coulomb information from integral generation to its final use.

The first check asks whether the tensor convention is complete.  A reported high-$l$ tensor should specify the chemists-to-pair mapping in \cref{eq:pair-convention}, the order of its four axes, the order of the real and complex spherical functions, the phase convention, and the normalization convention.  This descriptive information is often called metadata.  Without it, two codes can store the same physics in different index orders or with different signs, and a later transformation may silently combine incompatible tensors.

The second check measures local angular mixing.  For an isolated one-center shell, the total-$M$ rule is exact.  A molecular environment can move tensor weight outside those ideal blocks.  The benchmark should construct $V_{\mathrm{block}}$ with \cref{eq:block-projector} and report the off-block fraction $\eta_{\mathrm{off}}$ from \cref{eq:eta-off}.  It should also state how the local orbitals were chosen and made orthonormal, which center and axes were used, and whether the axes were rotated to reduce the mixing.  If a local orbital contains several angular momenta, the benchmark should report the fraction belonging to the shell being analyzed.  These details ensure that differences in $\eta_{\mathrm{off}}$ reflect the molecular system rather than hidden choices in the analysis.

The third check asks whether the angular blocks remain useful after factorization.  A Cholesky, RI, tensor-hypercontraction, or quantum low-rank calculation should report how many factors it produces and where their nonzero entries occur inside and outside the angular blocks.  It should also state the error threshold used to stop the decomposition, the rule used to choose each new factor, and the orbital basis in which the decomposition was performed.  These details matter because a reversible change of basis preserves exact rank, while the pattern of small numerical entries can change with the factorization procedure.

The fourth check asks whether the local tensor still has a compact physical interpretation.  For a correlated shell, the benchmark should report the full local tensor, the fitted Slater--Condon or related model parameters, and the residual, which is the part of the tensor that the fitted model does not reproduce.  Racah, Kanamori, Hubbard-$U$, or Hund-coupling values can then be given when the corresponding model is appropriate.  The residual shows how far the molecular interaction has moved away from the ideal rotationally symmetric shell.

The fifth check follows the labels to the end of the calculation.  When the same local block passes through a spinor transformation, active-space selection, low-rank factorization, or qubit mapping, the benchmark should record whether its angular and orbital labels remain traceable.  This final check connects integral generation to scientific interpretation because an efficient intermediate tensor has limited value if its later transformations cannot be reconstructed or verified.

Together, the five checks turn the representation argument into questions that can be answered with data.  Integral engines can report fully described pair objects.  Molecular calculations can measure angular mixing.  Factorization methods can show where their factors are nonzero at a stated accuracy.  Local-model builders can compare dense tensors with a few fitted physical parameters.  Spinor and quantum workflows can then verify that the same labels survive to their final outputs.  These results provide the evidence needed to guide the future interface described in the Outlook.

\section{Outlook}
\label{sec:outlook}

The most immediate opportunity is to make angular representation a usable part of the integral interface.  A real or complex solid-harmonic engine can return compact shell tensors together with explicit ordering, normalization, phase, and magnetic-quantum-number labels.  If these descriptors travel with the tensor, downstream codes can organize low-rank factors, local interactions, spinor transformations, and qubit mappings without reconstructing the angular meaning after integral generation.

Turning this idea into a practical standard will require benchmarks at two levels.  At the integral level, direct solid-harmonic and Cartesian-to-spherical routes should be compared with the same basis sets, thresholds, contraction patterns, and hardware.  At the tensor level, the resulting objects should be tested through $\eta_{\mathrm{off}}$, factor support, parameter-fit residuals, and convention checks.  Together, these measurements can reveal which benefits come from the smaller spherical representation, which arise from angular organization, and which depend on a particular implementation.

An especially useful next step is to map how one-center angular structure survives in molecules.  Total-$M$ blocks and compact Slater--Condon parameters provide exact reference patterns for an isolated rotationally invariant shell.  Across ligand fields, low-symmetry geometries, and heavy-element environments, controlled studies can determine when those patterns remain accurate enough to guide screening, factorization, or local-model construction.  Such studies would build a quantitative connection between atomic angular physics and molecular tensor algorithms.

Longer term, representation-aware ERIs could support closer integration across electronic-structure workflows.  Relativistic calculations could carry convention-controlled complex spherical blocks into spinor bases.  Local-Hamiltonian builders could fit and validate multiplet parameters directly from those blocks.  Quantum-simulation codes could test whether angular sectors improve factorization or reduce data movement before fermion-to-qubit mapping.  Because the same metadata would accompany each transformation, results would become easier to audit and compare across codes.

The broader outlook is therefore a shift from integral engines that return numerical arrays to integral engines that return structured Coulomb objects.  Such an interface would connect efficient integral generation to the symmetries, physical models, and computational choices that shape the rest of a many-electron calculation.

\section*{acknowledgments}
This work was supported by the Early Career Research Program of the U.S. Department of Energy, Office of Science, under Grant No. FWP 83466.

\bibliography{ref}

\end{document}